\documentclass[prl,twocolumn,showpacs,superscriptaddress,preprintnumbers,amssymb]{revtex4}
\usepackage{graphicx}
\usepackage{dcolumn}
\usepackage{bm}
\usepackage{wasysym}
\input epsf

\newcommand{\beq}{\begin{equation}}
\newcommand{\eeq}{\end{equation}}
\newcommand{\beqn}{\begin{eqnarray}}
\newcommand{\eeqn}{\end{eqnarray}}

\begin{document}
\title{Topological defects and the superfluid transition of the $s=1$ spinor condensate in two dimensions}
\author{Subroto Mukerjee}
\affiliation{Department of Physics, University of California,
Berkeley, CA 94720}
\affiliation{Materials Sciences Division,
Lawrence Berkeley National Laboratory, Berkeley, CA 94720}
\author{Cenke Xu}
\affiliation{Department of Physics, University of California,
Berkeley, CA 94720}
\author{J.~E.~Moore}
\affiliation{Department of Physics, University of California,
Berkeley, CA 94720} \affiliation{Materials Sciences Division,
Lawrence Berkeley National Laboratory, Berkeley, CA 94720}
\date{\today}
\begin{abstract}
The $s=1$ spinor Bose condensate at zero temperature supports ferromagnetic and polar phases that combine magnetic and superfluid ordering.  We analyze the topological defects of the polar condensate, correcting previous studies, and show that the polar condensate in two dimensions is unstable at any finite temperature; instead there is a nematic or paired superfluid phase with algebraic order in $\exp(2 i
\theta)$, where $\theta$ is the superfluid phase, and no magnetic order.  The Kosterlitz-Thouless transition out of this phase is driven by
unbinding of half-vortices (the spin-disordered version of the combined spin and phase defects found by Zhou), and the anomalous universal $8
T_c / \pi$ stiffness jump at the transition is confirmed in numerical simulations.  The anomalous stiffness jump is a clear experimental
signature of this phase and the corresponding phase transition.
\end{abstract}


\maketitle

Bose condensates of atoms with non-zero total spin, such as $^{87}$Rb~\cite{wieman2} and $^{23}$Na~\cite{gorlitz}, have recently been the focus
of intense experimental and theoretical study. The hyperfine degree of freedom in these systems allows complex ordered states,
distinct from those occurring in more conventional systems of spinless bosons, and combining superfluidity with different types of magnetic behavior. The dynamics and
topological defects of such states can be observed either destructively or {\it in situ}~\cite{stamperkurnreview}\cite{stamperkurn}. In
particular, the experimental study of two-dimensional condensates has been of special interest recently for at least three reasons: the
superfluid transition in two dimensions (2D) is of unconventional Kosterlitz-Thouless (KT) type~\cite{kosterlitzthouless} (i.e., driven by
unbinding of vortex defects); two-dimensional superfluids have power-law correlations of the quantum phase, rather than true long-range order;
and two-dimensional models are appropriate for some current experiments~\cite{stamperkurn}\cite{hadzibabic1}.

Topological defects and ordered phases in the $s=1$ (i.e., {\it total} spin $F=1$) spinor condensate have been discussed theoretically in many papers since the work of Ho~\cite{ho} and Ohmi and Machida~\cite{ohmimachida}. Experimentally,
$s=1$ systems are realized using atoms of $^{23}$Na, $^{39}$K and $^{87}$Rb with nuclear spin $I=3/2$.  In this letter, we resolve the nature of the
topological defects in the polar or antiferromagnetic phase by explicitly obtaining the order parameter manifold and its first homotopy group, then show that a new phase results in any polar $s=1$ condensate at finite temperature in 2D.  For the polar phase, Zhou~\cite{zhou} found the order parameter manifold $(U(1) \times S^2) / \mathbb{Z}_2$ by noticing an $\mathbb{Z}_2$ symmetry omitted in earlier work of Ho~\cite{ho}, but obtained an incorrect first homotopy group~\cite{zhou2}; that this homotopy group was incorrect was shown by Mak\"el\"a et al.~\cite{makela}, but these authors claimed that Zhou's order parameter manifold was also wrong.  We show that Zhou's order parameter manifold is correct but has the homotopy groups obtained indirectly in Ref.~\onlinecite{makela}.

These topological defects are crucial because they create a new phase in two dimensions: we find that the 2D polar condensate is unstable to a paired or nematic phase at any finite temperature.  This phase has algebraic order in $2 \theta$ where $\theta$ is the superfluid phase, no spin order, and a Kosterlitz-Thouless
superfluid transition driven by unconventional topological defects (half-vortices).  It has a clear experimental signature: an anomalous superfluid stiffness jump, which can be observed in an optically trapped condensate by the approach used by Hadzibabic {\em et.
al}~\cite{hadzibabic} to measure the conventional stiffness jump.

The Hamiltonian for the $s=1$ spinor condensate is~\cite{ho}
\begin{eqnarray}
H &=& \int d {\bf r} \Big(\frac{\hbar^2}{2M} \nabla \psi_a^+ \cdot
\nabla \psi_a + U({\bf r}) \psi_a^+ \psi_a \cr
&&+ \frac{c_0}{2}\psi_a^+ \psi_b^+
\psi_b \psi_a + \frac{c_2}{2} \psi_a^+ \psi_{a'}^+ {\bf
F}_{ab} \cdot {\bf F}_{a'b'}\psi_{b'}\psi_b \Big)
\label{hamiltonian}
\end{eqnarray}
where $a,b,a^\prime,b^\prime$ run from $m_z=+1$ to $m_z=-1$, $M$ is the mass of the bosons, $U({\bf r})$ is the trapping potential, and $c_2$
and $c_0$ are constants that depend on the strengths of the singlet and triplet scattering amplitudes. The matrices $\bf{F}$ are the $SU(2)$
generators in the $s=1$ representation.

The mean-field ground states for $U=0$ are homogeneous and unfragmented in the thermodynamic limit and can be decomposed using $\psi = \sqrt{n_0} \zeta$, where $\zeta$ is
a normalized spin-1 spinor~\cite{ho}\footnote{We strictly assume only that an ordered state with this decomposition exists on the basis of
  which a Ginzburg-Landau theory can be developed.}.  For $c_2 > 0$ the ground state is ``polar'', and for $c_2 < 0$ the ground state is ``ferromagnetic''; any polar state spinor
$\zeta_P$ and any ferromagnetic state spinor $\zeta_F$ can be obtained from
simple reference states using a phase $\theta$ and a rotation matrix $U$ in
the $s=1$ representation of $SU(2)$:
\beq
\zeta_P = e^{i \theta} U
\left( \begin{array}{c} 0 \\ 1 \\ 0 \end{array}\right), \quad \zeta_F = e^{i
  \theta} U \left( \begin{array}{c} 1 \\ 0 \\ 0 \end{array}\right).
\eeq The decomposition into phase $\theta$
and rotation matrix $U$ is not unique.

Topological defects in spinor condensate phases can be understood from homotopy groups of the order-parameter manifold~\cite{merminrev}. The
order-parameter manifold in the isotropic polar case has been previously studied: the original paper of Ho~\cite{ho} found a $U(1) \times S_2$ manifold,
where the $U(1)$ (sometimes written as $S^1$) denotes the manifold of values of the superfluid phase
$\theta$, and the $S^2$ refers to directions ${\bf \hat n}$ of the magnetic axis.  Later work~\cite{zhou} pointed out an additional ${\mathbb
Z}_2$ symmetry $(\theta \rightarrow \theta+\pi, {\bf \hat n} \rightarrow -{\bf \hat n})$ leading to the order-parameter manifold $M=(U(1) \times S^2) / \mathbb{Z}_2$
with defects that combine a half-vortex of the phase with a rotation from ${\bf \hat n}$ to $-{\bf \hat n}$.

We now derive $M$ from basic principles, showing that the subsequent claim of a different manifold~\cite{makela} is incorrect, but we find that the previously obtained homotopy groups~\cite{zhou2} of $M$ are incorrect and that the actual homotopy groups are those of Ref.~\onlinecite{makela}.
The key result, for readers interested in the nematic
superfluid phase, is the existence of phase half-vortices.  The overall symmetry group of an $s=1$ condensate is $G = SO(3)_S \times U(1)_G$, where the subscripts indicate
spin and gauge (phase) degrees of freedom. First consider the simpler ferromagnetic case: all states in this manifold can be obtained by a
symmetry operation $G$ on the reference state $(1,0,0)$. This state is left invariant by the group of operations (the residual group)
corresponding to an $SO(3)$ rotation by an angle $\alpha$ along the chosen $z$ axis and an overall phase shift of $-\alpha$ ($\alpha \in [0,
2\pi)$). The residual group is thus $H=U(1)$ and the order parameter manifold is $G/H \equiv SO(3)$.

The manifold $SO(3)$ for the ferromagnetic case also appears in the dipole-locked A phase of $^3$He, and the topological defects are well
understood~\cite{merminrev}.  Consider the polar case: for the polar reference state $(0,1,0)$, the residual symmetry group $H$ has two
disconnected parts: rotations about $z$, and rotations about $z$ followed by a rotation about $x$ (or $y$) by $\pi$ and multiplication by $-1$. The
resulting order parameter manifold $G/H$ can be written abstractly as~\cite{makela} $M^\prime = (SO(3)_S \times U(1)_G) / O(2)_{S+G}$, since $H \cong O(2)$. However, points on $M^\prime$ can be represented by a combination of a direction vector ${\bf \hat n}$
on $S^2$ and angle $\theta$ with $({\bf \hat n}, \theta)$ and $(-{\bf \hat n},\theta+\pi)$ identified, so $M^\prime = M.$

We write this identification as $\mathbb{Z}^{G+S}_2$ to stress that it links the gauge and spin symmetries: the resulting order parameter manifold is
\beq
M = {U(1) \times S^2 \over \mathbb{Z}^{G+S}_2} \not = U(1) \times {S^2 \over \mathbb{Z}_2} = M_n.
\eeq
That $\pi_1$ for the polar
state must equal $\mathbb{Z}$ was shown indirectly before~\cite{makela}, but it was also claimed that $M^\prime \not = M$; the form for $M$ is now used to prove explicitly that $\pi_1(M)=\mathbb{Z}$, not $\mathbb{Z} \times \mathbb{Z}_2 = \pi_1(M_n)$ as claimed in~\cite{zhou2}, and discuss the energetics of vortices.

Consider a closed path in $M$.  The starting point in $M$ is covered by two points of $M_2 = U(1) \times S^2$: pick one of these points and
trace out the preimage in $M_2$.  Either this path returns to the same point in $M_2$ or to the other identified point.  In the first case, the
path is identified with an element of $\pi_1(M_2) = \mathbb{Z}$: it wraps an integer number of times around the $U(1)$ part.  In the second
case, the path wraps a half-integer number of times around the $U(1)$ part, and moves from the initial point on the sphere to the antipodal
point.  Identify these paths with the 2-tuple $(n,e)$ in the first case, where $n$ is an integer, and $(n+\frac{1}{2},g)$ in the second case,
where $g^2=e$.  Concatenation of paths is then equivalent to addition in the first component, and $\mathbb{Z}_2$ multiplication in the second,
but the second entry is redundant and $\pi_1(M) = \mathbb{Z}$.

Note that although $\pi_1(M) = \mathbb{Z}$, just as for usual superfluid vortices, the fundamental vortex in $M$ combines both half a superfluid
vortex and a magnetic inversion. Odd vortex number in $\pi_1(M)$ means that the superfluid phase winds around the vortex by an odd multiple of
$\pi$, while the magnetic axis goes from ${\bf \hat n}$ to $-{\bf \hat n}$.  Only the vortices with even vortex number survive when the
anisotropy field favors $m_z=0$: these are ordinary vortices of the $m_z=0$ component.  Since $U(1) \times S^2$ is a covering space of $M$, all
higher homotopy groups are direct products $\pi_n(M) = \pi_n(U(1)) \times \pi_n(S^2)$: $\pi_2(M) = \pi_3(M) = \mathbb{Z}$, so in 3D there are
line defects, point defects, and coreless structures~\cite{shankar}.

The energies of the combined spin-superfluid defects described by the odd elements of $\pi_1(M)$ can be compared with those of normal superfluid
vortices at $T=0$ using (Eq.~\ref{hamiltonian}).  We ignore the finite vortex core energy in favor of the energy of the surroundings,
which is log-divergent in 2D.  The only contribution comes from the $|\nabla \Psi|^2$ term, and using the explicit
spinors~\cite{ho} gives, for stiffness $K$, \beq {\cal L} \approx \int\,d^2x\,{K \over 2} \left[ (\nabla {\bf \hat n})^2 + (\nabla \theta)^2
\right]. \eeq  The energy for the ordinary superfluid vortex is \beq E = \int_a^\infty \pi K r (2 \pi / 2 \pi r)^2 = \pi K \log(L/a), \eeq where
$L$ is the system size and $a$ the core size.  For the anomalous (combined) vortex, the energy cost is only half as large: there are now two
angular variables involved, but each runs over an angle $\pi$ rather than $2 \pi$.

A starting point to treat thermal fluctuations to study the physics in 2D is the nonlinear sigma model (NL$\sigma$M) on the order parameter
manifold. For the ordinary superfluid, this manifold is $U(1)$: the manifold is flat and the coupling is constant under renormalization-group
(RG) transformations, which underlies the existence of a Gaussian phase and finite-temperature KT transition~\cite{kosterlitzthouless}.  The
ferromagnetic manifold $SO(3) = S^3 / \mathbb{Z}_2$ is locally identical to $S^3$, which is curved in all three directions and flows to
weak coupling~\cite{polyakov}: no order is expected at $T>0$.

The polar state manifold $(S^2 \times U(1)) / \mathbb{Z}_2$ is more interesting: locally it is the same as $S^2 \times U(1)$, which has a
Kosterlitz-Thouless transition for the $U(1)$ phase but no $S^2$ order at finite temperature.  The renormalization-group flows (i.e., $\beta$-functions)
to all orders are unaffected by the $\mathbb{Z}_2$ identification, but the vortices in the superfluid phase now carry {\it half} the normal quantum of
vorticity.  Since the NL$\sigma$M analysis neglects amplitude fluctuations, it is desirable to check that such a phase actually exists: a prediction of
the above analysis of the isotropic polar phase is that the finite-temperature KT transition should be mediated by half-vortices of the phase.

The algebraically ordered state below the KT transition has algebraic correlations of $e^{2 i \theta}$, not $e^{i \theta}$, as $\theta$ is only
defined modulo $\pi$. Note that this finite-temperature sigma-model analysis for both phase and spin is distinct from the zero-temperature
analysis of spin with quantum fluctuations~\cite{zhou,wvliu} and from spin nematic phases of bosons in optical
lattices~\cite{zhoudemler}: the state we find at finite temperature is an algebraically ordered nematic superfluid with no spin order
and gapless excitations.  A transition mediated by half-vortices is also found for spinless bosons near a Feshbach resonance~\cite{radzihovsky}.

To picture this state, suppose that $e^{2 i \theta}$ had an expectation value rather than algebraic order: locally each component $\psi_\alpha$
averages to zero, but the spin-singlet combination $\psi_0 \psi_0 - 2 \psi_{+1} \psi_{-1} = \rho e^{2 i \theta} \not = 0$ and the total current can be nonzero: \beq {\bf
j}_s = {\hbar \over 2 i M} {\psi^*_\alpha \nabla \psi_\alpha} = {\hbar \rho \nabla \theta \over M}. \eeq Such a state can be thought of as
nematic, since order is only present in $2 \theta$, or paired, since the order appears in a two-boson operator if $\psi$ represents individual
bosons.

\begin{figure}[h!]
$\begin{array}{c}
\epsfxsize = 3in \epsfysize = 2in \epsffile{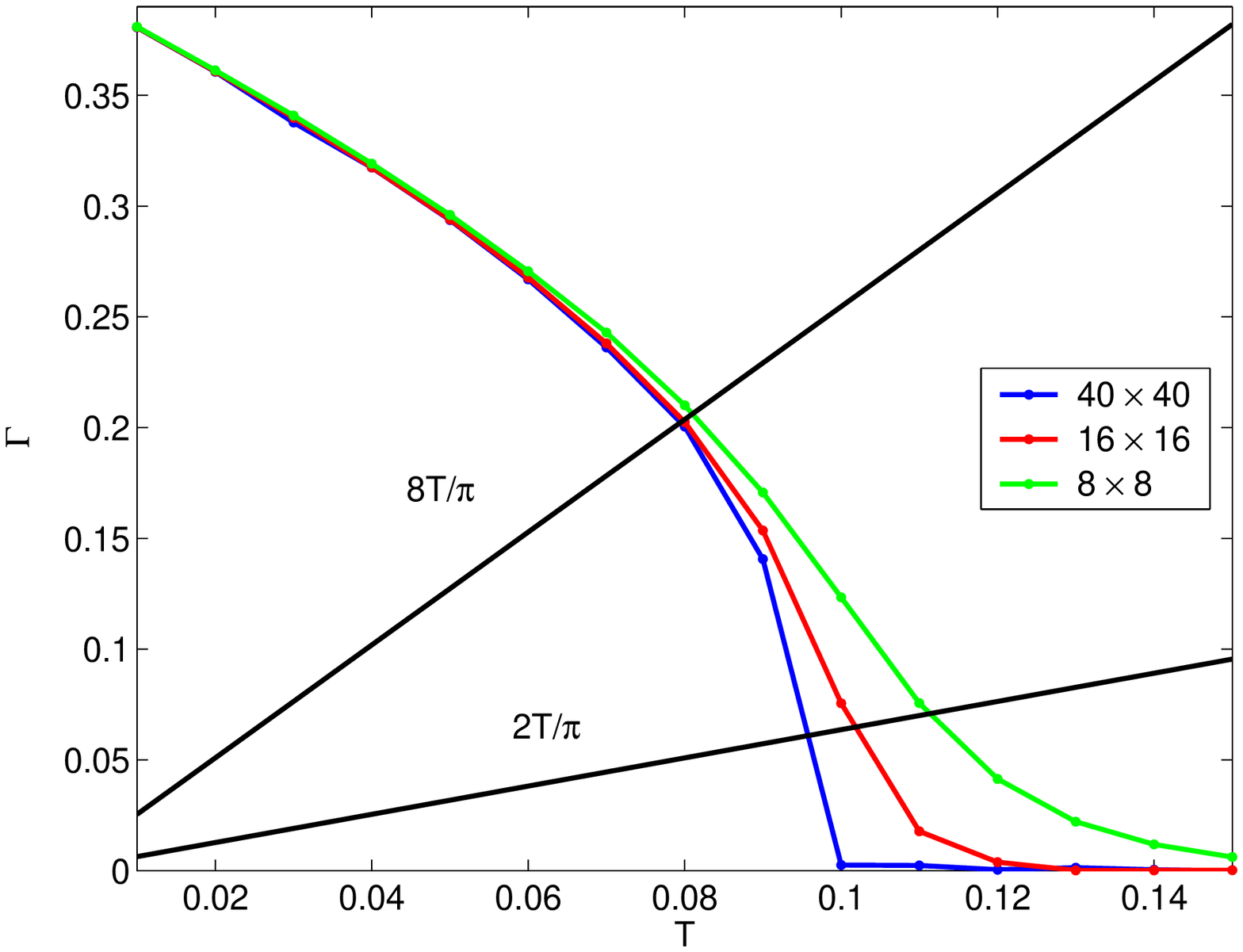} \\
\epsfxsize = 3in \epsfysize = 2in \epsffile{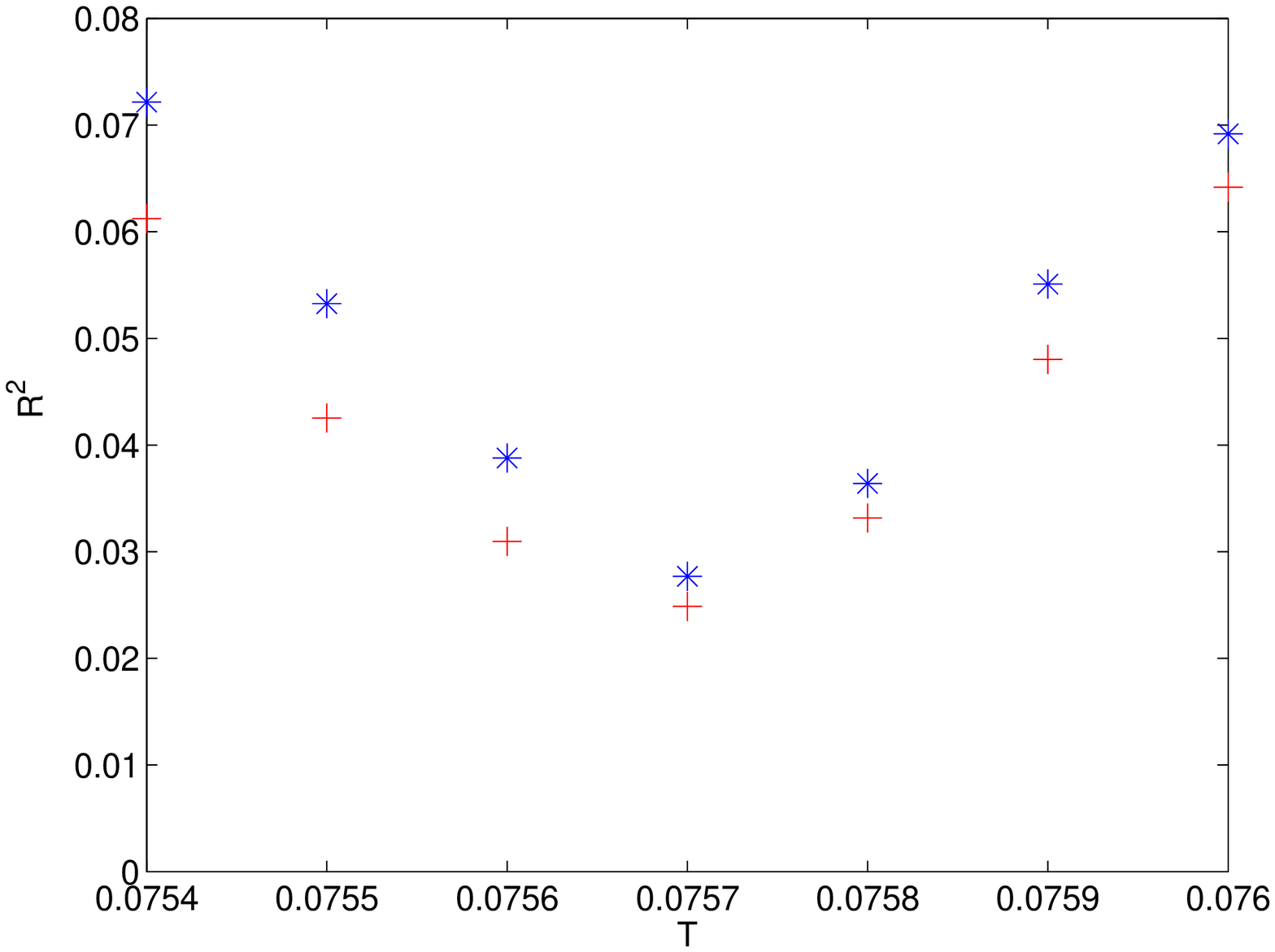}
\end{array}$
\caption{(Top) Helicity modulus $\Gamma$ as a function of temperature for a set of fixed values of the parameters, $\alpha=0.5$, $a_0=4.0$,
$T^{MF}_c=1.0$, $c_0=10.0$, $c_2=3.0$, $g_2=0$ (see Eq. 10) for different
system sizes. The two lines are $8T/\pi$ are $2T/\pi$. (Bottom) The goodness
of fit $R^2$ to Eq. 12 for different sets of system sizes as a function of
temperature. The asterisks and crosses correspond to sets with system sizes 6,
8, 10, 12, 14, 16, 18, 20 and 10, 12, 14, 16, 18, 20, 24, 32 respectively.}
\end{figure}

A direct check on the above scenario is that the Kosterlitz-Thouless transition will occur when half-vortices unbind.  The KT jump in the renormalized
stiffness or ``helicity modulus'' at the transition must therefore be four times larger than the conventional value:
\beq
\rho_c = 4 \rho_c^0 = {8 T_c \over \pi}.
\label{bigjump}
\eeq
The superfluid stiffness jump by $\rho_c^0$ in superfluid helium was observed by Bishop and Reppy~\cite{bishopreppy}.  Recently the KT transition of a
spinless atomic BEC was observed~\cite{hadzibabic}: the prediction that an isotropic $s=1$ polar condensate will have a jump 4 times as large is a
direct experimental test for the superfluid nematic phase.

Numerical Monte Carlo simulations of the polar phase (Fig. 1) reveal a jump in the helicity modulus compatible with the enhanced value
(Eq.~\ref{bigjump}) in the thermodynamic limit, and clearly distinct from the conventional jump $\rho^c_0$.  An additional check is provided by
turning on spin-space anisotropy. The complete phase diagram with spin-space anisotropy
introduced through the term \beq H^\prime = \int d{\bf r} g_2\,(\psi^\dagger_{+1} \psi_{+1} + \psi^\dagger_{-1} \psi_{-1}) \eeq is shown in
Fig. 2. 
\begin{figure}[h!]
\epsfxsize = 3.0in
\epsffile{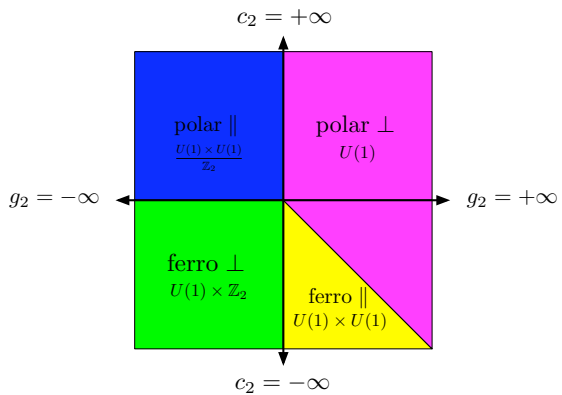} \caption{The phase diagram of the ground-state of the $F=1$ spinor
condensate with uniaxial anisotropy (Eq. 3).  The order parameter manifold is shown for each phase.}
\end{figure}
With $m_z=0$ favored ($g_2 > 0$), an ordinary KT transition occurs, and is
observed numerically.  With $m_z=\pm 1$ favored ($g_2 < 0$), the
order-parameter manifold is reduced to $(U(1) \times U(1)) / \mathbb{Z}_2$, where half-vortices of the in-plane spin are bound to half-vortices
of the phase, and for weak anisotropy the anomalous jump survives. The experimental
source of this anisotropy is the quadratic
Zeeman coupling to an external field.

The numerics use the Ginzburg-Landau free energy
\begin{widetext}
\begin{equation}
F = \int d {\bf r} \left[\alpha (\nabla \psi_a)^* (\nabla \psi_a) + a_0(T-T^{MF}_c)\psi^*_a \psi_a + \frac{c_0}{2}\psi_a^* \psi_b^* \psi_b
\psi_a + \frac{c_2}{2} \psi_a^* \psi_{a'}^* {\bf F}_{ab} \cdot {\bf
  F}_{a'b'}\psi_{b'}\psi_b + g_2\,(\psi^\dagger_{+1} \psi_{+1} + \psi^\dagger_{-1} \psi_{-1})\right],
\end{equation}
\end{widetext}
in which the KT transition temperature ($T_{KT}$) is renormalized below $T^{MF}_c$ by the quartic terms. The transition is identified via the jump in the helicity modulus $\Gamma$, defined by introducing a twist $\phi$ in the phase $\theta$ in (2):
\begin{equation}
\Gamma = \left(\left \langle \frac{\partial^2 F}{\partial \phi^2}\right \rangle - \frac{1}{k_BT}\left \langle \left(
\frac{\partial F}{\partial \phi}\right)^2 \right \rangle \right)\Bigg|_{\phi = 0}
\end{equation}
Metropolis Monte Carlo simulations were performed for square lattices with sizes varying from $6 \times 6$ to
$64 \times 64$. $10^9$ Monte Carlo steps were used at each value of temperature for equilibration and calculation of the helicity modulus.

The plot of $\Gamma$ as a function of temperature $T$, for $c_2 > 0$ and
$g_2=0$ (no anisotropy) for different system sizes $N$ is shown in Fig. 1. A jump in $\Gamma$ is clearly visible for the larger systems. The two lines have slopes $8T/\pi$ and $2T/\pi$, and the jump is near the intersection of the former with the $\Gamma$ curves. To establish a KT transition and locate $T_c$, the finite-size scaling of $\Gamma(N,T)$ is investigated. The same procedure was used for the KT transition in the
XY model~\cite{minnhagen}. The helicity modulus scales as $N \rightarrow \infty$ as
\begin{equation}
\Gamma(N,T_c) = \Gamma_\infty \left[1 + \frac{1}{2} \frac{1}{\ln N + C} \right].
\label{fsfit}
\end{equation}
Here $\Gamma_\infty$ ($=8T_c/\pi$ in this case) is the value of the jump in the helicity modulus in the thermodynamic limit and $C$ is a
constant. Even though Eq.~\ref{fsfit} is strictly valid only for large values of $N$, it is found to hold even for small values in the case of
the regular KT transition. $\Gamma (N,T)$ for different values of $N$ and $T$ is fit to the expression Eq.~\ref{fsfit} with $C$ as an adjustable
parameter. The goodness of fit $R^2$ for different sets of system sizes ${N}$ as a function of temperature $T$ is plotted in Fig. 2. There is a
minimum in the value of $R^2$ at a temperature $T = 0.0757$, which is identified with $T_{KT}$. The value $T_{KT}$ is robust against the choice
of different sets of system sizes ${N}$ for the fit. A similar procedure was carried out with a value of $\Gamma_\infty = 2T_c/\pi$: no such
minimum was found to exist in the vicinity of a corresponding normal KT transition. This offers strong evidence of the existence of an anomalous
KT transition and the nematic superfluid phase.

To conclude, we have obtained the order parameter manifold and topological defects of the polar $s=1$ spinor condensate.  Using a NL$\sigma$M
analysis of this order parameter manifold together with numerical Monte-Carlo simulations, we have demonstrated the existence of a
low temperature spin-disordered phase with quasi-long-range phase order.  The KT transition out of this phase, driven by unbinding of half-vortex defects, can be detected experimentally by its anomalous $8T_c/\pi$ stiffness jump.

The authors wish to acknowledge conversations with D. Podolsky, D. M. Stamper-Kurn, and A. Vishwanath, and support from DOE (S. M.), NSF DMR-0238760
(C. X. and J. E. M.), and the IBM SUR program.

\bibliography{atomfinbib}
\end{document}